# Recoil proton polarization: A new discriminative observable for deeply virtual Compton scattering

Olga Bessidskaia Bylund, Maxime Defurne, and Pierre A. M. Guichon

*Irfu, CEA, Université Paris-Saclay, 91191 Gif-sur-Yvette, France*

Generalized parton distributions describe the correlations between the longitudinal momentum and the transverse position of quarks and gluons in a nucleon. They can be constrained by measuring photon leptoproduction observables, arising from the interference between Bethe-Heitler and deeply virtual Compton scattering (DVCS) processes. At leading-twist/leading-order, the amplitude of the latter is parametrized by complex integrals of the GPDs $\{H, E, \tilde{H}, \tilde{E}\}$. As data collected on an unpolarized or longitudinally polarized target constrains $H$ and $\tilde{H}$, $E$ is poorly known as it requires data collected with a transversely polarized target, which is very challenging to implement in fixed-target experiments. The only alternative considered so far has been DVCS on a neutron with a deuterium target, while assuming isospin symmetry and absence of final-state interactions. Today, we introduce the polarization of the recoil proton as a new DVCS observable, highly sensitive to $E$, which appears feasible for an experimental study at a high-luminosity facility such as Jefferson Lab.

## I. INTRODUCTION

Obtaining a complete understanding of the strong interaction requires an accurate description of the behavior of the quarks and gluons confined in a nucleon. This behavior is encoded by a set of structure functions. For instance, the longitudinal momentum distribution is described by parton distribution functions (PDFs), while transverse momentum distributions (TMDs) give correlations between the transverse and longitudinal momentum of partons inside a nucleon. Finally, generalized parton distributions (GPDs) encode correlations between the transverse position and the longitudinal momentum of partons [1–5].

From the first principles of quantum chromodynamics (QCD), theorists strive to develop nonperturbative approaches [6,7] to compute PDFs, TMDs or GPDs. Their predictions must be tested against experimental measurements to validate the theoretical approach. Whereas stringent constraints on PDFs and TMDs have been provided by 40 years of measurements of lepton-proton deep-inelastic cross sections, Drell-Yan data and other processes, GPDs are not as well known, mostly due to the small cross-sections of GPD-sensitive leptoproduction processes. The factorization theorem states that the amplitude of such processes can be written as a convolution between the GPDs and a kernel, derived using perturbative QCD to describe the absorption of the virtual photon by a quark and the emission of a photon or a meson. For a single high-energy photon emission by the quark, also known as deeply virtual Compton scattering (DVCS) as displayed in Fig. 1, the factorization theorem has been proven valid at all twists and all orders in perturbation theory. DVCS has been considered a golden channel for GPD studies, not only because of the well-established validity of the factorization theorem, but also because of the accessibility to the phase of the DVCS amplitude, a unique feature among leptoproduction processes, which is allowed by the interference term with the Bethe-Heitler process. As depicted by Fig. 1, the Bethe-Heitler process has the same final state as DVCS, but differs from the latter by the photon being radiated by the incoming or the outgoing lepton.

At leading-twist, the DVCS amplitude is parameterized by four Compton form factors $\{\mathcal{H}, \mathcal{E}, \tilde{\mathcal{H}}, \tilde{\mathcal{E}}\}$, each being the convolution of the corresponding chiral-even GPD $\{H, E, \tilde{H}, \tilde{E}\}$ with the hard kernel. Measurements with an un- or longitudinally polarized target are respectively mostly sensitive to $\mathcal{H}$ and $\tilde{\mathcal{H}}$. As such target polarizations are fairly easy to implement experimentally, data collected with HERMES [8], COMPASS [9,10], CLAS [11–17], and Hall A [18–21] of Jefferson Lab provide fair constraints of these two CFFs.

To obtain information on $\mathcal{E}$ through DVCS, a transversely polarized target (TPT) is typically required. In a collider, it is very feasible to provide a transversally polarized proton beam, so good constraints are expected from the future electron-ion collider, reaching a luminosity high enough to study DVCS in the small-$x_B$ region. However, regarding the valence region probed with fixed-target experiments, maintaining the transverse polarization of a target under an intense electron beam is a considerable challenge. So far, only the HERMES collaboration provided DVCS data using a gaseous TPT [22,23], which implied a greatly reduced luminosity compared to unpolarized liquid targets. As an alternative to TPT, the neutron offers a better access to $\mathcal{E}$ than the proton as its electromagnetic form factors enhance the contribution of $\mathcal{E}$ in the BH/DVCS interference term. Assuming no initial/final state interaction between the recoil neutron and the spectator proton of the nucleus, deuterium targets were employed as a neutron target at Jefferson Lab [24,25]. However, aside from the previously mentioned assumptions, the neutron detection efficiency tends to reduce the effective luminosity. Finally, DVCS is poorly sensitive to $\tilde{\mathcal{E}}$.

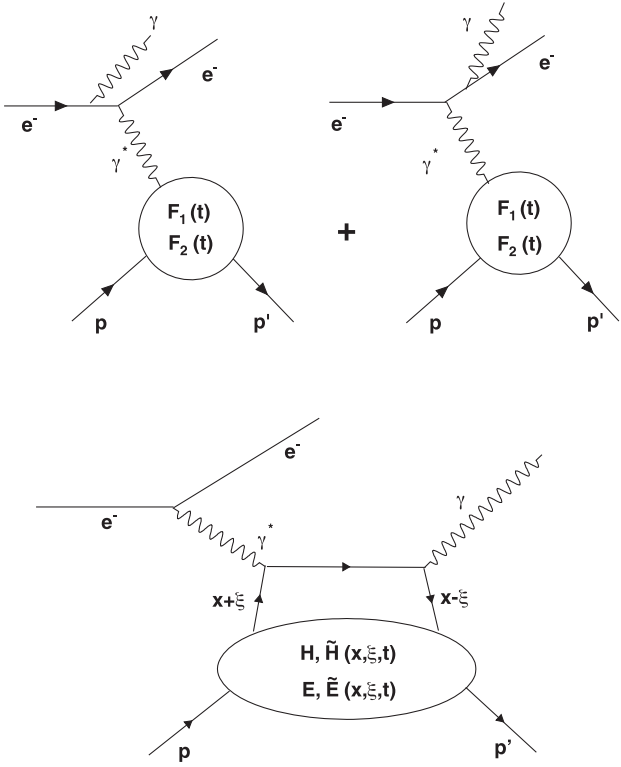

FIG. 1. Diagrams of Bethe-Heitler (top) and DVCS (bottom). For Bethe-Heitler, the nucleon structure is parameterized by the form factors depending on t = $|p' - p|^2$. In DVCS, the GPDs parametrize the nucleon structure. In addition to t, they depend on $x$ and $\xi$ being respectively the average longitudinal momentum carried by the active quark and half the longitudinal momentum transfer to the proton.

From the DVCS data, phenomenologists have tested several approaches to extract most or all CFFs [26–29]. But the CFF-$\mathcal{E}$ remains poorly known as strong experimental constraints are missing for the later.

In this article, we study the polarization of the recoil proton in photon electroproduction as a new experimental observable. This observable appears to be sensitive to $\mathcal{E}$ using an unpolarized target and is suggested as an alternative to a measurement with a transversely polarized target, with the main benefit of enabling higher luminosities. First the theoretical computation of the formulas relating the recoil proton polarization to the Compton form factors are derived in Sec. II. Further, the polarization and its sensitivity to the various CFFs are studied around a kinematic point of interest, using the Kroll-Goloskokov GPD model [30]. Then in Sec. III we proceed to show that an experiment appears to be feasible and would strongly discriminate between the predictions of various models. For the basic principles of proton polarimetry we refer to Appendix.

## II. THEORY

We consider the photon electroproduction $(e, e'\gamma)$ reaction:

$$e(k, h_e) + N(p, h) \to e(k', h'_e) + N(p', h') + \gamma(q', \lambda') \quad (1)$$

where the letters in parenthesis are the momenta and helicities of the particles. The amplitude $T^{ee'\gamma}$ for this reaction is the sum of the Bethe-Heitler (BH) and virtual Compton scattering (VCS) amplitudes

$$T^{ee'\gamma} = T_{\text{BH}} + T_{\text{VCS}}. \quad (2)$$

In the following we shall denote by $q = k - k'$ the momentum of the virtual photon in the VCS amplitude. The center of mass (c.m.) system will be the one defined by $\vec{p} + \vec{q} = \vec{p}' + \vec{q}' = 0$. We compute $\mathcal{M}$ in the c.m. frame and below every quantity refers to this frame, until specified otherwise. We choose a system of axis $\mathcal{R}$ such that the contravariant components of the 4-vectors are

$$q = (q^0, 0, 0, |\vec{q}|)$$
$$p = \left(\sqrt{M^2 + \vec{q}^2}, 0, 0, -|\vec{q}|\right)$$
$$q' = (q'^0, q'^0 \sin\theta, 0, q'^0 \cos\theta).$$

In this coordinate system the azimuthal angle of the lepton plane is $\phi_h$. As usual we note

$$Q^2 = -q^2, \quad x_B = \frac{Q^2}{2p \cdot q},$$
$$s = (p + q)^2 = M^2 + Q^2 \frac{1 - x_B}{x_B}, \quad t = (p - p')^2. \quad (3)$$

We limit our considerations to the cross section and the recoil polarization with a polarized beam. The target is not polarized and the final photon helicity is not detected. The observables are deduced from the following quantity (the factor 1/2 is just the density matrix of the unpolarized target):

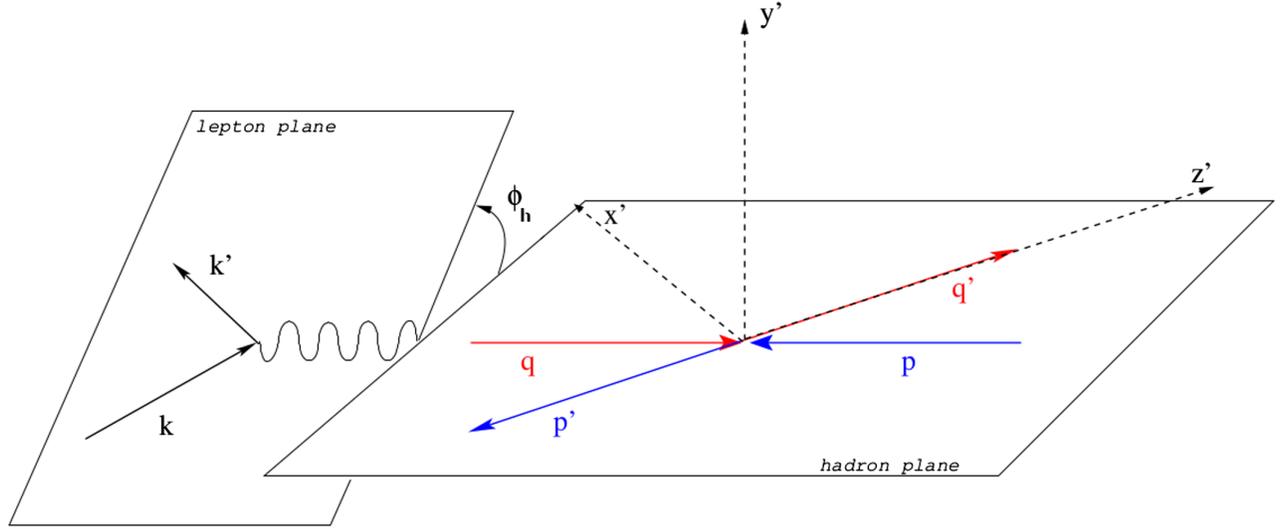

FIG. 2. The coordinate system in which $P'$ is computed. The center of mass system of the hadronic plane, spanned by the directions of the recoil proton and the produced photon, is used. The $z'$-axis is along the direction of the produced photon. The $y'$-axis is normal to the hadronic plane, its direction given by the cross product of the directions of the virtual photon and the produced photon. The $x'$-axis is in the hadronic plane, forming a right-handed coordinate system with the two other axes. The angle between the hadronic plane and the leptonic plane, spanned by the directions of the incident electron and the final electron, is denoted $\phi_h$.

$$\mathcal{M}(\vec{a}) = \sum_{h',\bar{h}'} e(h',\bar{h}') \sum_{h,h_e,h'_e,\lambda'} T^{ee'\gamma}(\bar{h}',h'_e,\lambda';h,\bar{h}_e) \frac{1}{2}\rho_e(\bar{h}_e,h_e) T^{ee'\gamma*}(h',h'_e,\lambda';h,h_e), \quad (4)$$

with $\vec{a}$ an auxiliary 3-vector, which refers to the rest frame of the recoiling proton, and where the beam density matrix is (neglecting the transverse polarization, the effect of which is suppressed by the electron mass):

$$\rho_e(\bar{h}_e, h_e) = \delta(\bar{h}_e, h_e)\frac{1 + h_e P_{e\|}}{2}. \quad (5)$$

Here $P_{e\|}$ is the longitudinal polarization of the lepton. The efficiency matrix $e$ is defined as

$$e(\bar{h}, h) = \delta(\bar{h}, h)(1 - ha_z) + \delta(\bar{h}, -h)(-a_x + iha_y).$$

One has

$$\mathcal{M}(\vec{a}) = \mathcal{M}(0) + \sum_{i=x,y,z} \mathcal{M}(i)a_i.$$

The Lorentz invariant cross section is

$$d\sigma = e^6 d\Gamma \mathcal{M}(0) \quad (6)$$

$$d\Gamma = \frac{(2\pi)^{-5} ds dQ^2 dt d\phi_h d\phi_e}{128(p.k)^2 \sqrt{4sQ^2 + (s - M^2 - Q^2)^2}}, \quad (7)$$

where $e = \sqrt{4\pi\alpha}$ and $\phi_e$ is the (trivial) angle of the lepton plane in the lab.

In order to describe the polarization of the recoil proton, the coordinate system $\mathcal{R}$ is rotated such that the new $z'$-axis is along the direction of the produced photon and hence antiparallel with the direction of the recoil proton in the CM frame, see Fig. 2. Let $P' = (P'_x, P'_y, P'_z)$ be the rest frame components of the recoiling proton polarization in this coordinate system.

From the parametrization of the recoiling proton density matrix:

$$\rho'_p(\bar{h}, h) = \delta(\bar{h}, h)\frac{1 - hP'_z}{2} + \delta(\bar{h}, -h)\frac{-P'_x + ihP'_y}{2}$$

and

$$\rho'_p(\bar{h}', h') = \mathcal{N}^{-1} \sum_{h,h_e,h'_e,\lambda'} T^{ee'\gamma}(\bar{h}', h'_e, \lambda'; h, \bar{h}_e)$$

$$\times \frac{1}{2}\rho_e(\bar{h}_e, h_e) T^{ee'\gamma*}(h', h'_e, \lambda'; h, h_e). \quad (8)$$

Here the normalization $\mathcal{N}$ is determined by the condition $\sum_{h'} \rho'_p(h', h') = 1$ and we get the result

$$P'_i = \frac{\mathcal{M}(i)}{\mathcal{M}(0)}. \quad (9)$$

The BH amplitude has the standard expression [31] and the VCS amplitude is

$$T_{\text{VCS}} = \frac{Q_{\text{Beam}}}{Q^2} \epsilon'^{*\mu}_{\lambda'} H_{\mu\nu} \bar{u}(k', h'_e) \gamma^\nu u(k, h_e)$$

with

$$H_{\mu\nu} = -i \int d^4x e^{-iq\cdot x} \langle p'h'|T[J_\nu(x)J_\mu(0)]|ph\rangle,$$

where $J$ is the conserved hadronic current (with unit charge). One defines the CM helicity amplitudes (see Ref. [31] for the polarization vectors)

$$M(\lambda'h', \lambda h) = \epsilon'^{*\mu}_{\lambda'} H_{\mu\nu} \epsilon^\nu_\lambda,$$

which are functions of $s, t, Q^2$. Within our phase conventions, reflexion symmetry with respect to the scattering plane implies

$$M(\lambda'h', \lambda h) = (-)^{\lambda+\lambda'+(h-h')/2} M(-\lambda' - h', -\lambda - h).$$

Therefore the reaction depends in general on 12 independent complex amplitudes. In the twist-2 approximation that we use below the photon helicity is conserved, which brings the number of independent amplitudes to 4. They are related to the four twist-2 Compton form factors (CFF) $\mathcal{H}, \mathcal{E}, \tilde{\mathcal{H}}, \tilde{\mathcal{E}}$ in the formalism of [32].

As $P'$ is defined for convenience with respect to the CM-frame, it is not the polarization vector that would be measured by a proton polarimeter. A final set of boosts and rotations are applied to get $P^m$—the polarization in the rest frame of the recoil proton with the $z^m$-axis collinear to the proton momentum in the lab frame and the $y^m$-axis orthogonal to the hadronic plane.

To study the polarization sensitivity to the CFFs, we use primarily the Goloshokov-Kroll (GK) model for the generalized parton distributions $H, E, \tilde{H}, \tilde{E}$. The implementation is done with the PARTONS software [33].

The sensitivity of the polarization to the CFFs is a function of the beam energy $E_k$, $Q^2$, $x_B$, $t$ and $\phi_h$. Apprehending it in five dimensions and providing a concise description is complicated. For the sake of clarity in this section, we focus on a local kinematic study of the polarization around $Q^2 = 1.8$ GeV$^2$, $x_B = 0.17$, $t = -0.45$ GeV$^2$, $\phi_h = 180°$ with a beam energy of 10.6 GeV and beam helicity $h_e = 1$. This choice results from a search of the best kinematic settings to experimentally constrain CFF-$\mathcal{E}$ over the five dimensions maximizing:

(i) the sensitivity of the polarization to $\mathcal{E}$,
(ii) the performance of a polarimeter targeting a scattered proton with its momentum given by the DVCS kinematics. The polarimeter performance is a function of the proton momentum and is related to the polarization measurement accuracy, as summarized in Appendix,
(iii) the expected number of produced DVCS recoil protons, which is related to the photon electroproduction cross section.

TABLE I. Compton form factors from the GK-model at $Q^2 = 1.8$ GeV$^2$, $x_B = 0.17$, $t = -0.45$ GeV$^2$.

| $\mathcal{H}$ | $\mathcal{E}$ | $\tilde{\mathcal{H}}$ | $\tilde{\mathcal{E}}$ |
|---|---|---|---|
| $-1.1 + 5.4i$ | $-2.4 - 0.4i$ | $0.7 + 1.8i$ | $38 + 19i$ |

Table I summarizes the CFF values used for the plots and discussions. We are currently considering a beam polarization of 100%, as $P^m_x$ and $P^m_z$ are 0 with an unpolarized beam.

In Figs. 3–6, the $E_k/Q^2/x_B/t$-dependences of the three components of the polarization are shown. The polarization is smoothly and gradually changing across the studied ranges, except when $t$ is near $t_{\min}$. An additional study, consisting of setting one CFF to 0 at a time to see how it affects all three components of the polarization, is performed to assess their sensitivity. The difference between the nominal prediction and the prediction where one of the CFFs is set to 0 is referred to as $\Delta P^m_i$. It appears that $P^m_x$ depends mostly on $\tilde{\mathcal{H}}$ and $\tilde{\mathcal{E}}$, $P^m_y$ on $\mathcal{H}, \mathcal{E}$ and $\tilde{\mathcal{E}}$, while $P^m_z$ is not showing much sensitivity to any CFF except possibly $\mathcal{H}$. Focusing on $P^m_y$ and $\mathcal{E}$, one clearly notices the increasing sensitivity with increasing $E_k$ and $|t|$ and with decreasing $Q^2$. The contribution from $\mathcal{E}$ only has a weak dependence on $x_B$. Whereas these kinematics were optimized to access $\mathcal{E}$ through $P^m_y$, it seems that the approach would have given a similar result for $P^m_x$ with $\tilde{\mathcal{H}}$.

Regarding the $\phi_h$-dependence, the polarization is first decomposed into BH, DVCS and interference contributions, displayed by Fig. 7, keeping only the associated amplitude at a time in $\mathcal{M}^m(x/y/z)$ of Eq. (9), while keeping all contributions for $\mathcal{M}^m(0)$ in the denominator —thus the three contributions add to give the total polarization. The beam-helicity independent ($P^u$) and dependent ($P^h$) parts of the polarization are defined as following:

$$P^m_{x/z} = h_e(P^u_{x/z} + h_e P^h_{x/z}), \quad (10)$$

$$P^m_y = P^u_y + h_e P^h_y. \quad (11)$$

Averaging over the beam helicity, one obtains $P^u_y$ and $P^h_{x/z}$. On the contrary, $P^h_y$ and $P^u_{x/z}$ are isolated by making the difference between the helicity state of the beam. Unlike with $E_k$, $Q^2$, $x_B$ and $t$, all polarization components exhibit a nontrivial sizable $\phi_h$-dependence. $P^u_x$ and $P^u_z$ are both dominated by the BH contribution at small $\phi_h$, while approaching 180°, the three terms are of similar magnitude. Note that the DVCS contribution is much larger for $P_x^u$ than $P_z^u$, most likely indicating a better sensitivity to CFFs with $P_x^u$. Considering $P_y^u$, BH is inducing almost no polarization, but it is still diluting the polarization signal at small $\phi_h$, where its contribution to $\mathcal{M}(0)$ is much larger than that of the other two terms. A large polarization is induced in the range 90°–270°, particularly by the interference term.

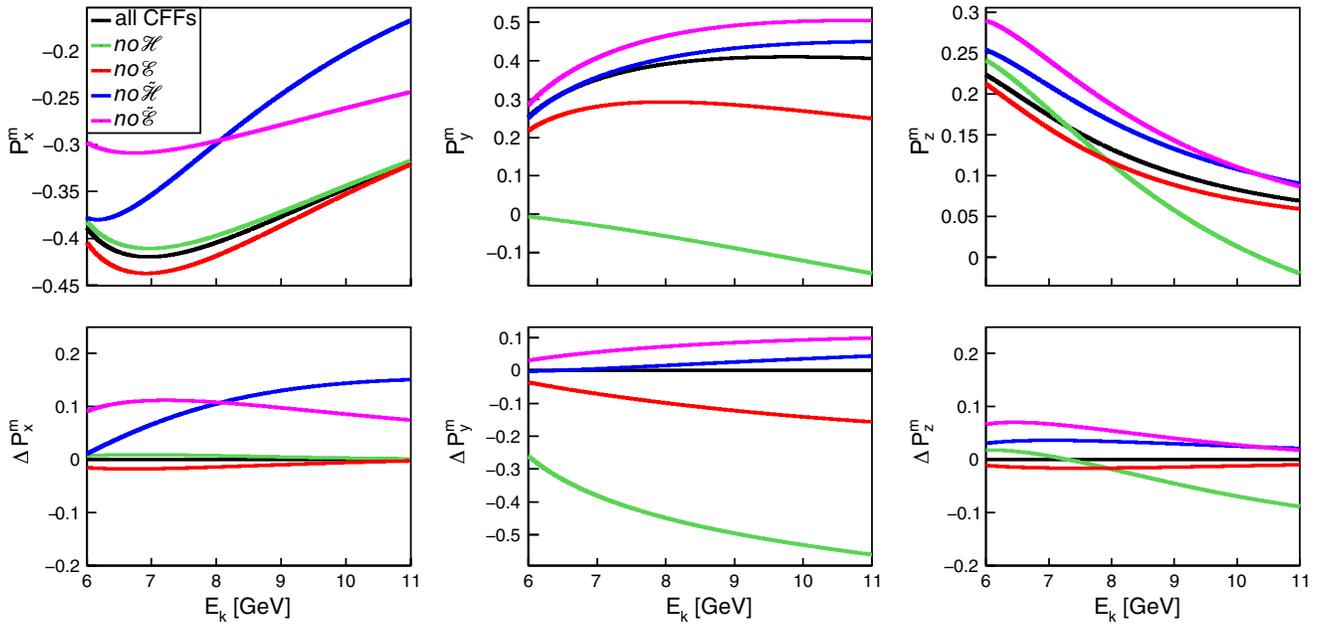

FIG. 3. Top: the polarization along the x-axis (left), y-axis (middle), and z-axis (right) as a function of beam energy for the configuration $Q^2 = 1.8$ GeV$^2$, $x_B = 0.17$, $t = -0.45$ GeV$^2$, $\phi_h = 180°$. The curves show the prediction from the GK model (plain black with $h_e = 1$), and with one of the coefficients at a time switched off. Bottom: the difference between the polarization using all CFFs and the polarization setting a CFF to 0.

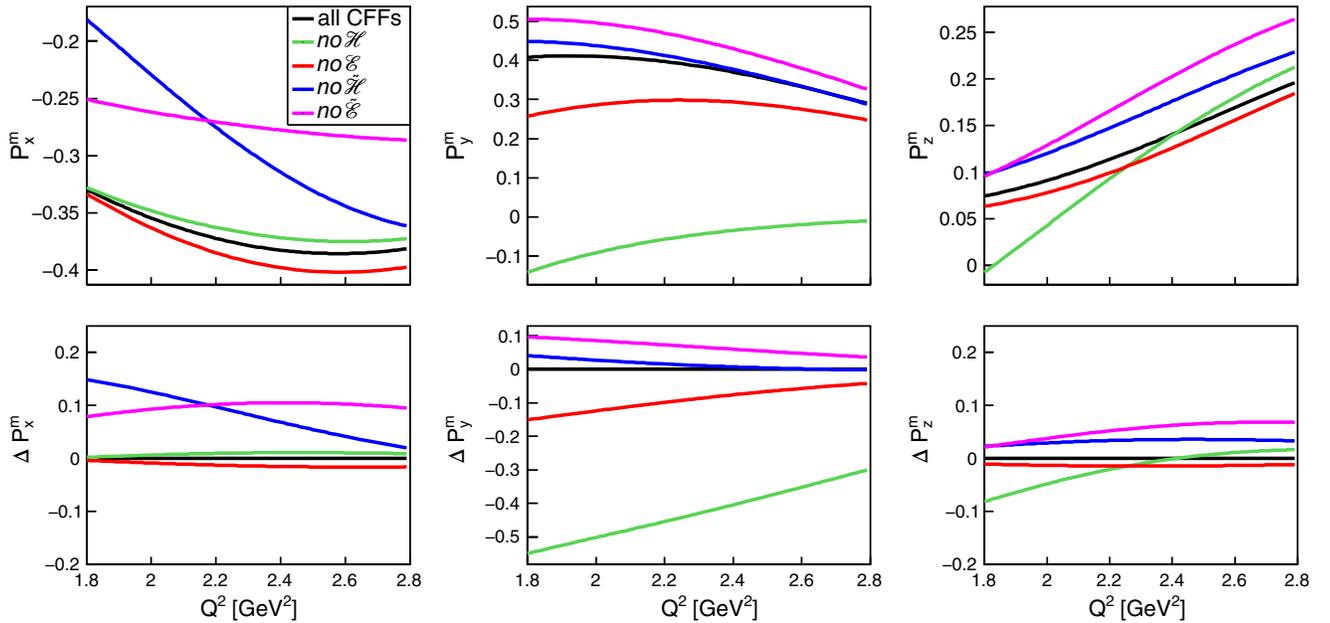

FIG. 4. Top: the polarization along the x-axis (left), y-axis (middle) and z-axis (right) as a function of $Q^2$ for the configuration $E_k = 10.6$ GeV, $x_B = 0.17$, $t = -0.45$ GeV$^2$, $\phi_h = 180°$. The curves show the prediction from the GK model (plain black with $h_e = 1$), and with one of the coefficients at a time switched off. Bottom: the difference between the polarization using all CFFs and the polarization setting a CFF to 0.

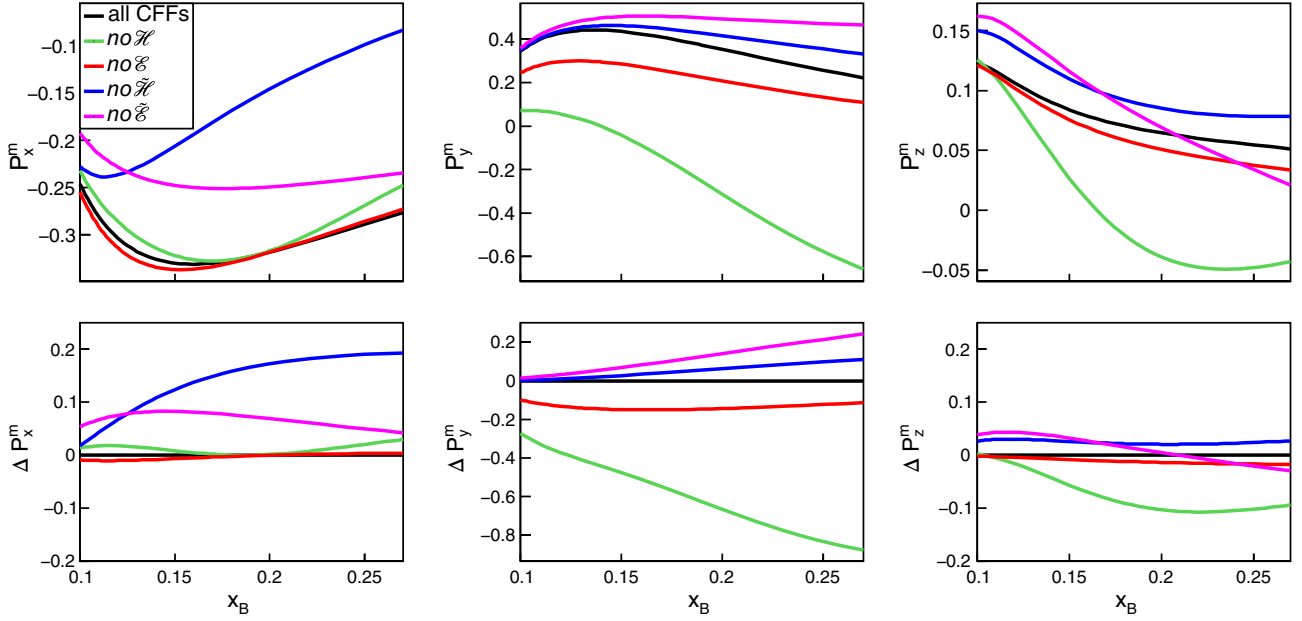

FIG. 5. Top: the polarization along the x-axis (left), y-axis (middle) and z-axis (right) as a function of $x_B$ for the configuration $E_k = 10.6$ GeV, $Q^2 = 1.8$ GeV$^2$, $t = -0.45$ GeV$^2$, $\phi_h = 180°$. The curves show the prediction from the GK model (plain black with $h_e = 1$), and with one of the coefficients at a time switched off. Bottom: the difference between the polarization using all CFFs and the polarization setting a CFF to 0.

Regarding the beam-helicity dependent terms $P^h_{x/y/z}$, the induced polarization arises mostly from the interference term, which is largest at 70° and 290° for $P^h_{x/z}$ and at 90° and 270° for $P^h_y$. For x and z-components, $P^h$ is as large as $P^u$. Finally, all beam-helicity dependent terms of the polarization vanish at $\phi_h = 0$ or 180°, as they seem to be carried by $\sin(\phi_h)$ and $\sin(2\phi_h)$-harmonics.

In Fig. 8, the polarization components are shown as a function of $\phi_h$ with $h_e = 1$ to study the CFF-dependence

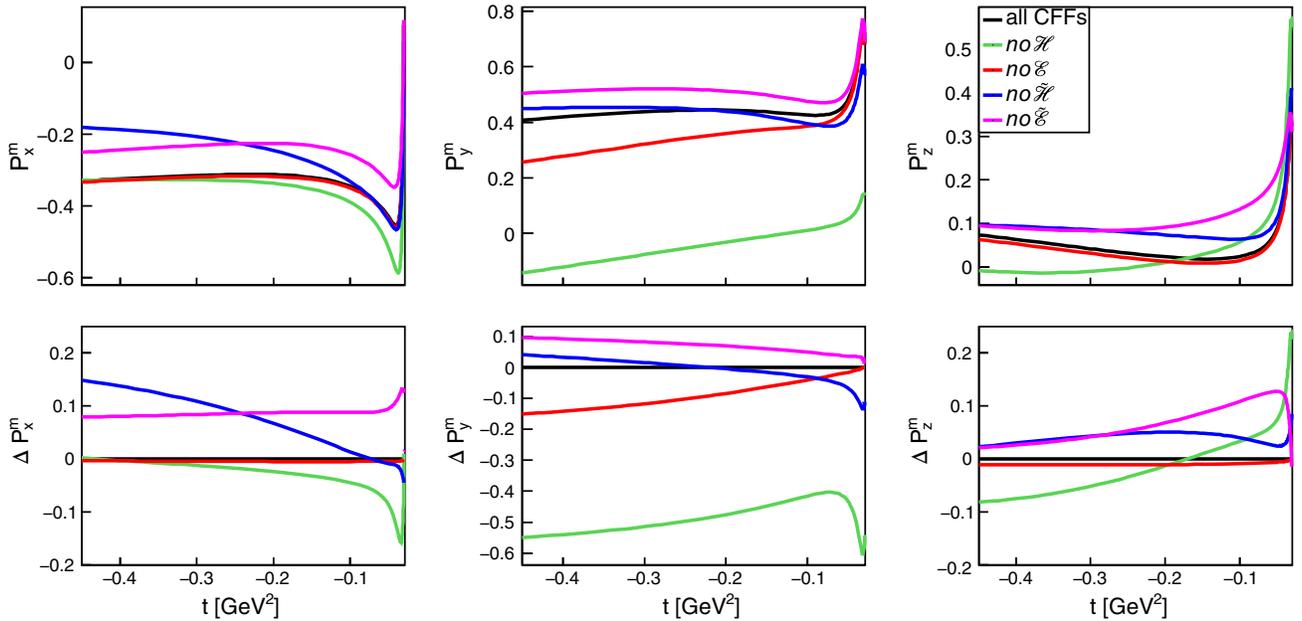

FIG. 6. Top: The polarization along the x-axis (left), y-axis (middle) and z-axis (right) as a function of $t$ for the configuration $E_k = 10.6$ GeV, $Q^2 = 1.8$ GeV$^2$, $x_B = 0.17$, $\phi_h = 180°$. The curves show the prediction from the GK model (plain black with $h_e = 1$), and with one of the coefficients at a time switched off. Bottom: The difference between the polarization using all CFFs and the polarization setting a CFF to 0.

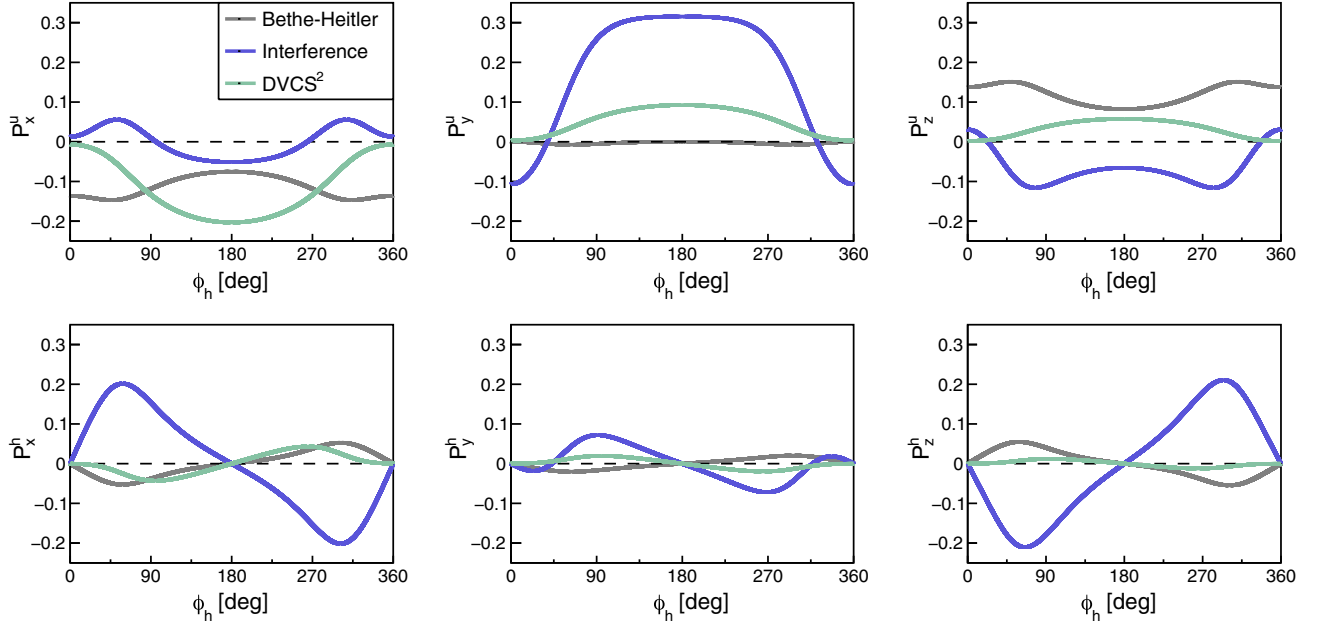

FIG. 7. Separation of the BH, interference and DVCS contributions to the recoil proton polarization for the configuration $E_k = 10.6$ GeV, $Q^2 = 1.8$ GeV$^2$, $x_B = 0.17$ and $t = -0.45$ GeV$^2$. The top row displays the beam-helicity independent polarization ($P^u$) while bottom row shows the beam-helicity dependent terms ($P^h$) as a function of $\phi_h$.

by alternatively setting them to 0. In other words, the plain black curve is given by $P_i^u + P_i^h$ from Fig. 7 for $i \in [x, y, z]$. At $\phi_h = 0$ the effect of the polarization is small for all three components as the polarization is mostly driven by the Bethe-Heitler here. Looking at CFF-$\mathcal{E}$, $P_y^m$ seems to change the most when setting $E$ to 0 around $\phi_h = 180°$, while $P_x^m$ and $P_z^m$ are almost completely insensitive to $E$ over the entire $\phi_h$-range. $P_y^m$ also appears to have a high sensitivity to $\tilde{\mathcal{H}}$ when studied this way. When setting $\tilde{\mathcal{E}}$ to 0, all three components are changing by quite much, as seen also with the other kinematical variables. This feature can be explained by the large size of the CFF-$\tilde{\mathcal{E}}$ in the GK model

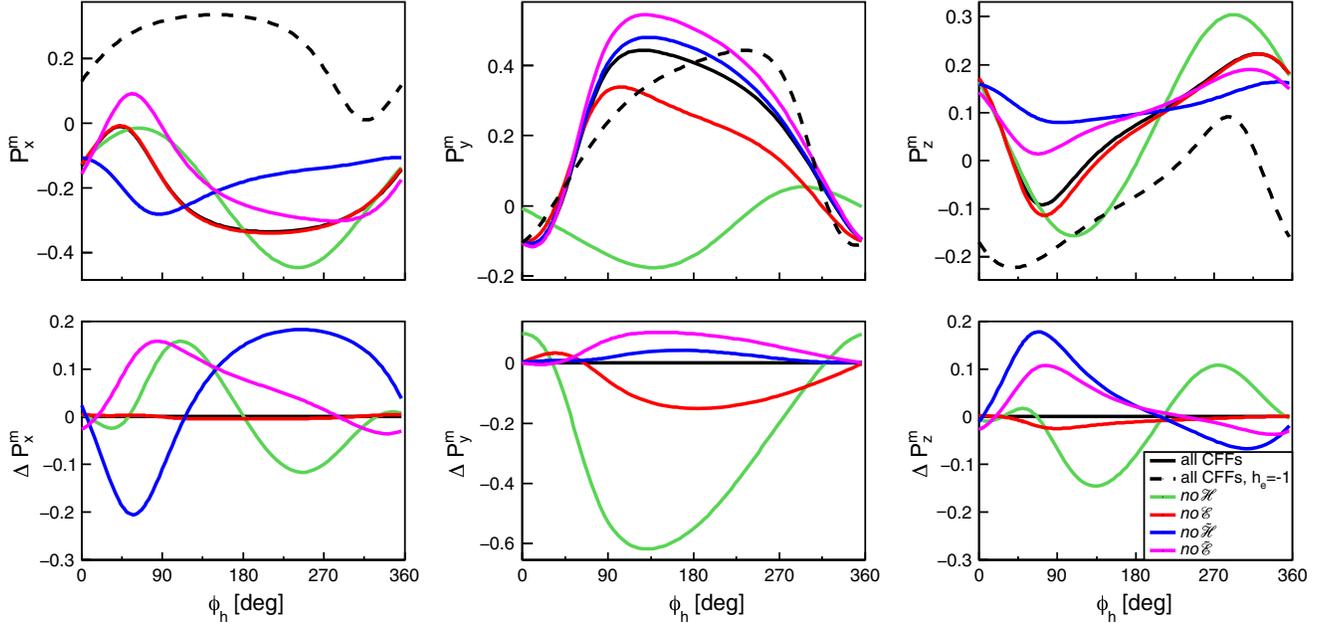

FIG. 8. Top: the polarization along the $x$-axis (left), $y$-axis (middle) and $z$-axis (right) as a function of $\phi_h$ for the configuration $E_k = 10.6$ GeV, $Q^2 = 1.8$ GeV$^2$, $x_B = 0.17$, $t = -0.45$ GeV$^2$. The curves show the prediction from the GK model (plain black with $h_e = 1$, dashed gray for $h_e = -1$) and with one of the coefficients at a time switched off. Bottom: the difference between the polarization using all CFFs and the polarization setting a CFF to 0.

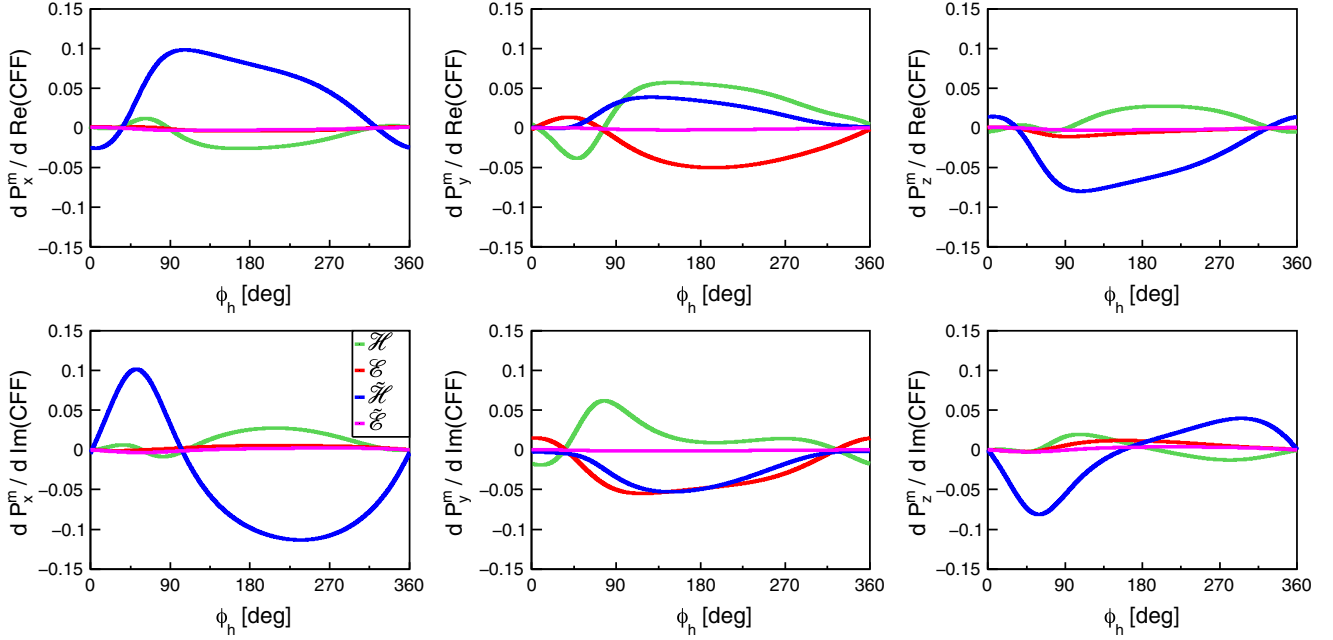

FIG. 9. The derivative of $P_x^m$ (left), $P_y^m$ (middle) and $P_z^m$ with respect to the real part (top) or imaginary part (bottom) of each CFF as a function of $\phi_h$ for $E_k = 10.6$ GeV, $Q^2 = 1.8$ GeV$^2$, $x_B = 0.17$, $t = -0.45$ GeV$^2$ and $h_e = 1$.

and cannot only be attributed to an intrinsic kinematical sensitivity of the polarization.

To get a more direct access to this kinematical sensitivity, the derivatives of the $P^m$-components with respect to the real or imaginary parts of the CFFs are computed and shown in Fig. 9. Indeed the derivative of the interference term, the latter being larger than the DVCS$^2$-contribution most of the time as seen in Fig. 7, reduces to the kinematical coefficient in front of the CFF. However, as the DVCS term is bilinear in CFFs, some model-dependence remains. As seen in Fig. 9, the derivatives related to $\tilde{\mathcal{E}}$ are very close to 0 and confirm that switching off a CFF is a model-biased approach to assess the sensitivity of the polarization. At $\phi_h = 180°$, $P_y^m$ is showing a better sensitivity to $\mathcal{E}$ than to $\mathcal{H}$ for the imaginary part in particular. The sensitivity to $\tilde{\mathcal{H}}$ is of similar magnitude, which is elaborated on further down. Regarding $P_x^m$, there is almost no sensitivity to $\tilde{\mathcal{E}}$ and $\mathcal{E}$, and it seems that $P_x^m$ is mostly sensitive to $\tilde{\mathcal{H}}$.

$$\mathcal{M}^m(x) = -20.42 + 19.06\,\mathrm{Re}\tilde{\mathcal{H}} + 7.15\,\mathrm{Re}\mathcal{H} - 1.04\,\mathrm{Re}\mathcal{E} - 0.56\,\mathrm{Re}\tilde{\mathcal{E}}$$
$$- 2.93(\mathcal{H}\tilde{\mathcal{H}}^* + \mathcal{H}^*\tilde{\mathcal{H}}) + 0.16(\mathcal{E}\tilde{\mathcal{H}}^* + \mathcal{E}^*\tilde{\mathcal{H}})$$
$$+ 0.04(\mathcal{H}\tilde{\mathcal{E}}^* + \mathcal{H}^*\tilde{\mathcal{E}}) + 0.03(\mathcal{E}\tilde{\mathcal{E}}^* + \mathcal{E}^*\tilde{\mathcal{E}}) \quad (12)$$

$$\mathcal{M}^m(y) = 15.50\,\mathrm{Im}\mathcal{H} - 10.05\,\mathrm{Im}\mathcal{E} + 3.44\,\mathrm{Im}\tilde{\mathcal{H}} - 0.44\,\mathrm{Im}\tilde{\mathcal{E}}$$
$$+ 1.51\,\mathrm{Im}(\mathcal{E}\mathcal{H}^* - \mathcal{E}^*\mathcal{H}) + 0.14\,\mathrm{Im}(\tilde{\mathcal{E}}\tilde{\mathcal{H}}^* - \tilde{\mathcal{E}}^*\tilde{\mathcal{H}}) \quad (13)$$

$$\mathcal{M}^m(z) = 22.24 - 12.56\,\mathrm{Re}\tilde{\mathcal{H}} - 2.93\,\mathrm{Re}\mathcal{H} - 2.37\,\mathrm{Re}\mathcal{E} - 0.48\,\mathrm{Re}\tilde{\mathcal{E}}$$
$$+ 0.98(\mathcal{E}\tilde{\mathcal{H}}^* + \tilde{\mathcal{H}}\mathcal{E}^*) + 0.40(\mathcal{H}\tilde{\mathcal{H}}^* + \tilde{\mathcal{H}}\mathcal{H}^*) + 0.09(\mathcal{H}\tilde{\mathcal{E}}^* + \tilde{\mathcal{E}}\mathcal{H}^*) - 0.01(\mathcal{E}\tilde{\mathcal{E}}^* + \tilde{\mathcal{E}}\mathcal{E}^*) \quad (14)$$

$$\mathcal{M}^m(0) = 72.55 - 29.61\,\mathrm{Re}\mathcal{H} - 13.25\,\mathrm{Re}\tilde{\mathcal{H}} - 4.93\,\mathrm{Re}\mathcal{E} + 0.01\,\mathrm{Re}\tilde{\mathcal{E}}$$
$$+ 4.51(\mathcal{H}\mathcal{H}^* + \tilde{\mathcal{H}}\tilde{\mathcal{H}}^*) + 0.51\mathcal{E}\mathcal{E}^*$$
$$- 0.04(\mathcal{H}\mathcal{E}^* + \mathcal{H}^*\mathcal{E} + \tilde{\mathcal{H}}\tilde{\mathcal{E}}^* + \tilde{\mathcal{H}}^*\tilde{\mathcal{E}}) + 0.005\tilde{\mathcal{E}}\tilde{\mathcal{E}}^* \quad (15)$$

To remove all dependence on the GK model in the assessment of the sensitivity, a local expression of $\mathcal{M}^m(i) = \mathcal{M}(0) \cdot P_i^m$ for $i \in \{x, y, z\}$ is derived at $\phi_h = 180°$, $E_k = 10.6$ GeV, $Q^2 = 1.8$ GeV$^2$, $x_B = 0.17$, $t = -0.45$ GeV$^2$, $h_e = 1$ and shown in Eqs. (12)–(15). The full general expression is far too lengthy and involved to present here. The constant terms in $\mathcal{M}^m(x)$, $\mathcal{M}^m(z)$ and $\mathcal{M}(0)$ are induced by the BH contribution. The terms that appear at $\phi_h = 180°$ are all helicity-independent, as anticipated from Fig. 7. The BH does not induce any polarization along y, as confirmed by the absence of a constant term in $\mathcal{M}^m(y)$. Linear terms are interference terms and bilinear terms are DVCS$^2$-terms. Comparing the kinematical coefficients in front of the linear or bilinear CFF-terms in $\mathcal{M}^m(x)$, $\mathcal{M}^m(y)$ and $\mathcal{M}^m(z)$ confirms that the conclusions drawn from the derivatives are not only features of the GPD model, but kinematically justified. We see that:

(i) $P_x^m$ is mostly sensitive to $\tilde{\mathcal{H}}$: the real part is accessible mainly through the DVCS/BH interference and the imaginary part from a DVCS$^2$-term combining both $\mathcal{H}$ and $\tilde{\mathcal{H}}$.

(ii) $P_y^m$ is sensitive to the imaginary part of $\mathcal{E}$ thanks to the DVCS/BH interference. Additionally, this sensitivity is enhanced by the DVCS-term entangling $\mathcal{E}$ with the well-known $\mathcal{H}$, which also provides sensitivity to the real part of $\mathcal{E}$. The apparent sensitivity of $P_y^m$ to $\tilde{\mathcal{H}}$ comes partly from a convolution of $\tilde{\mathcal{H}}$ with the unconstrained $\tilde{\mathcal{E}}$, which has large values in the GK model.

(iii) $P_z^m$ arises mostly from the BH/DVCS interference and from the large BH term, the latter of which makes it less interesting than $P_x^m$ to study CFFs.

### III. EXPERIMENTAL FEASIBILITY

In an experimental context, as the DVCS cross section is small and a rescattering of the recoil proton in an analyzer is required to measure the polarization (see Appendix), a high luminosity is definitely a prerequisite. Therefore, the Hall C of Jefferson Lab is the most suitable for a DVCS recoil proton polarization (RPP) fixed target experiment. For our study, we assume a beam current of 10 μA impinged on a 15-cm long liquid hydrogen target and a data-taking time of three weeks.

At Hall C, the high-momentum spectrometer (HMS) [34] would detect scattered electrons and a PbWO$_4$-based neutral particle spectrometer (NPS) could detect the DVCS photons, as described by the accepted proposal for DVCS cross section measurements in Hall C [35]. The DVCS Hall C GEANT4 simulation package [36] has been used to describe both the HMS and the NPS. For the polarimeter, a toy model, based on a 15-cm thick Carbon analyzer, was developed using McNaughton's parametrizations [37] to describe the analyzing power and Bonin et al.'s parametrization for the efficiency [38]. A polarimeter acceptance of $40° \times 60°$ in $\theta \times \phi$ for the proton scattering angles was assumed, corresponding to approximately one steradian of coverage, and an upper cut on proton momentum of 1.010 GeV/c is applied to exclude large $|t|$-events.

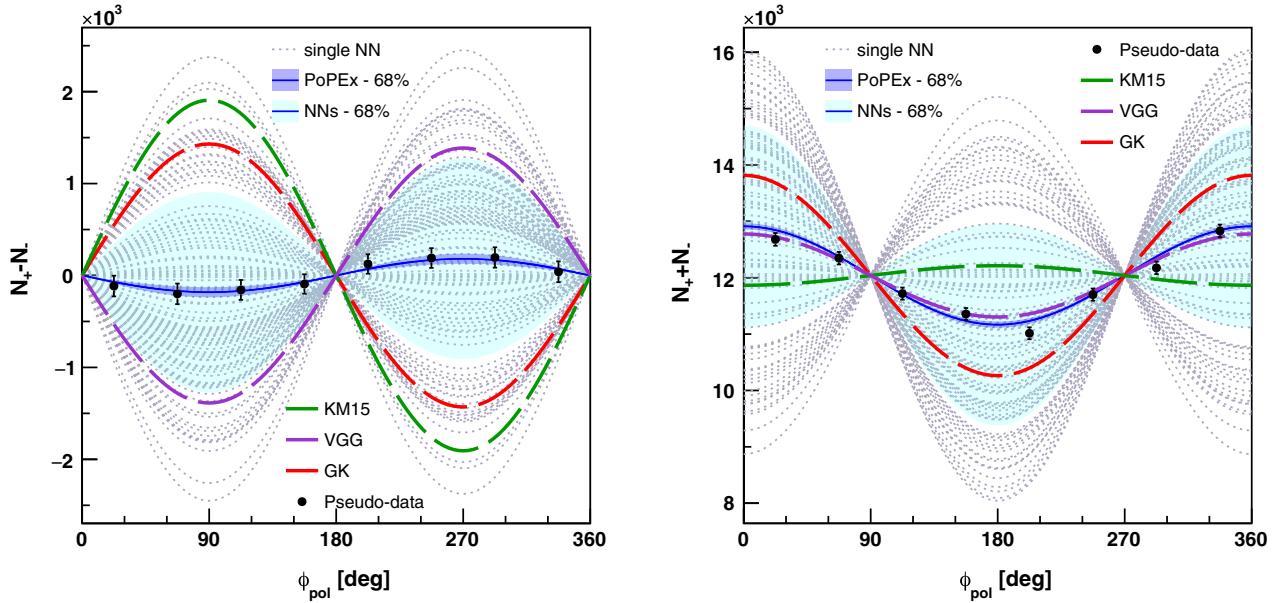

FIG. 10. The azimuthal rescattering angle $\phi_{pol}$ for $\theta_{pol}$ in the range 4°–19° when subtracting over helicities, which gives sensitivity to $P_x^m$ (left) and when adding all events, which gives sensitivity to $P_y^m$ (right). The dotted lines show the NN replicas, with the 68% confidence level band shown in turquoise. The pseudodata is shown in black dots and the fit to pseudodata (referred to as PoPEx) in blue, with the blue band illustrating the statistical uncertainty on the fit result. The GK, VGG and KM15 models are shown for comparison in red, purple and green respectively.

TABLE II. The predictions for the Compton form factors from the GK model, the VGG model and the KM15 model. The DVCS parameters at which the CFFs are evaluated are $Q^2 = 1.8$ GeV$^2$, $x_B = 0.17$, $t = -0.45$ GeV$^2$, $\phi_h = 180°$ and helicity 1.

| Prediction | $\mathcal{H}$ | $\mathcal{E}$ | $\tilde{\mathcal{H}}$ | $\tilde{\mathcal{E}}$ | $P_x^m$ | $P_y^m$ |
|---|---|---|---|---|---|---|
| GK | $-1.1 + 5.4i$ | $-2.4 - 0.4i$ | $0.7 + 1.8i$ | $38 + 19i$ | $-0.33$ | $0.41$ |
| VGG | $-2.2 + 4.8i$ | $-1.0 + 1.5i$ | $0.5 + 1.4i$ | $51$ | $0.32$ | $0.17$ |
| KM15 | $-2.9 + 3.2i$ | $1.6$ | $0.5 + 1.5i$ | $124$ | $-0.44$ | $-0.04$ |

As previously mentioned, great care has been taken in the choice of the kinematics, since the discriminative power of an experimental measurement of the RPP is the result of a convolution between the statistical power (related to the DVCS cross section), the polarimeter analyzing power and efficiency, as well as the polarization sensitivity to the targeted CFFs. As explained in the introduction, the extractions of $\mathcal{H}$ and $\tilde{\mathcal{H}}$ are already benefiting from data collected on an unpolarized or longitudinally polarized target. However, very little is known from $\mathcal{E}$ in the valence region, as it typically requires a transversely polarized target. Consequently, the best kinematics were determined to maximize the constraint set on $\mathcal{E}$. In addition, a few additional constraints were taken into account such as:

(i) $|t|/Q^2 \lesssim 0.25$ to limit higher-twist contributions,
(ii) particles momentum and direction compatible with detector acceptances (and an isolation of at least 10° between all particles).

Already introduced in the previous section, the following set of kinematics was thus found to be suitable: $E_k = 10.6$ GeV, $Q^2 = 1.8$ GeV$^2$, $x_B = 0.17$, $t = -0.45$ GeV$^2$, 10.6 GeV being the currently maximal beam energy available at Jefferson Lab.

To derive experimental predictions, the library composed of 101 artificial neural networks (ANNs) released by the PARTONS collaboration [27] has been used. These ANNs have been produced from a global fit of DVCS datasets including all JLab 6-GeV data except neutron ones: the average prediction of these ANNs is the fit prediction, while their dispersion represents the associated uncertainty. As the kinematics of interest is within the JLab 6-GeV phase space, the ANN predictions are relevant and will be used to estimate the statistical significance of a RPP measurement. In Fig. 10, the expected azimuthal distributions of the rescattered protons in the C-analyzer are shown by the black markers and the 68%-confidence level of this prediction by the ANNs is represented by the light blue band. Considering the statistical uncertainty $\sigma_P \sim 0.013$ of the azimuthal distributions displayed by the dark blue band, approximately 90% of the ANNs would be rejected by measurements of such precision with a $3\sigma_P$-cut. Such a measurement would also strongly discriminate between the GPD models for both $P_x^m$ and $P_y^m$, their predictions disagreeing by over 0.2 for $P_y^m$ and above 0.1 for $P_x^m$, as seen in Table II.

## IV. CONCLUSION

We report for the first time the computation of the recoil proton polarization for the DVCS process. Its sensitivity to CFF $\mathcal{E}$ was demonstrated and studied as an alternative to using a transversely polarized target, which is difficult to implement for a fixed-target experiment. It offers as well excellent perspectives to constrain also $\tilde{\mathcal{H}}$. Consequently, measuring the recoil proton polarization is equivalent to running with a target being simultaneously polarized longitudinally and transversely. A careful search of suitable kinematics for a dedicated experiment was performed. Assuming 3 weeks of beamtime at 10 μA with a polarimeter covering 1 sr with the DVCS Hall C experimental setup, it was found that the statistical power allows to easily discriminate between all considered GPD models and put stringent constraints on both CFFs $\mathcal{E}$ and $\tilde{\mathcal{H}}$.

While we have chosen a DVCS configuration sensitive to $\mathcal{E}$ using inputs from the GK model, we must keep in mind that the polarization depends on the other CFFs as well. For a more complete interpretation, a global analysis taking into account constraints on the other CFFs through a global fit should be performed to assess the impact of the polarization measurement.

We have at this stage only considered the statistical uncertainty. For the purpose of designing an experiment and submitting a proposal, the accidental background in the polarimeter must be taken into account. The exclusive $\pi^0$ background to DVCS would need to be accounted for as well in the polarization extraction. Although, if such an experiment is allocated some beam time, it would be perfectly suited as well to measure the proton polarization for $\pi^0$-electroproduction, and help in isolating the transversity GPD contributions as it does for DVCS with $E$ and $\tilde{H}$.

## ACKNOWLEDGMENTS

We would like to thank E. Tomasi-Gustafsson for suggestions and explanations regarding proton polarimetry, C. Munoz Camacho for help with using the GEANT4 simulation for Hall C, H. Moutarde and P. Sznajder for discussions of input models for the GPDs and N. d'Hose, F. Bossù and F. Sabatié for helpful comments and advice. This work was funded by the Programme Exploratoire of commissariat a l'énergie atomique et aux énergies alternatives.

## APPENDIX: POLARIMETRY

Our objective is to connect the observable recoil proton polarization to the Compton form factors. Here we provide a summary of the general concepts of proton polarimetry relevant for our study.

Due to the spin-orbit coupling, when transversally polarized protons strike a nucleus $N$, an azimuthal dependence in the $pN$-cross section will be induced. Exploiting this effect, one can measure the average polarization of a set of protons in a statistical measurement. A proton polarimeter is therefore constituted of an analyzer off which the protons will scatter, surrounded by a set of trackers upstream and downstream the analyzer to determine the polar $\theta_{\text{pol}}$ and azimuthal $\phi_{\text{pol}}$ scattering angles in the polarimeter.

To define $\theta_{\text{pol}}$ and $\phi_{\text{pol}}$, we are following the convention found in [39] with x- and y-directions arbitrary as long as they are orthogonal to each other and to the incoming proton. With $\mathbf{p}'$ being the incoming proton momentum and $\mathbf{p}_{\text{pol}}$ the rescattered proton momentum, the normal vector $\hat{\mathbf{n}}$ is defined as:

$$\hat{\mathbf{n}} = \mathbf{p}' \times \mathbf{p}_{\text{pol}}/|\mathbf{p}' \times \mathbf{p}_{\text{pol}}| \quad (A1)$$

The scattering angles are then given by

$$\sin\theta_{\text{pol}} = |\mathbf{p}' \times \mathbf{p}_{\text{pol}}|,$$
$$\sin\phi_{\text{pol}} = -\hat{\mathbf{x}} \cdot \hat{\mathbf{n}}, \qquad \cos\phi_{\text{pol}} = \hat{\mathbf{y}} \cdot \hat{\mathbf{n}}. \quad (A2)$$

The following distribution is then observed:

$$\frac{dN}{d\theta_{\text{pol}}} = N_0 \cdot \frac{d\epsilon}{d\theta_{\text{pol}}} \cdot (1 + A_p(P_y \cos\phi_{\text{pol}} - P_x \sin\phi_{\text{pol}})). \quad (A3)$$

with $N_0$ is the total number of incident protons, $\frac{d\epsilon}{d\theta_{\text{pol}}}$ and $A_p$ the differential efficiency and the analyzing power of the polarimeter and $P_x$, $P_y$ being the transverse polarizations to the proton momentum. The analyzing power gives the sensitivity of the scattering process to the proton polarization. In this work we are considering a carbon analyzer with density 1.7 g/cm$^3$.

Both the efficiency and the analyzing power depend on the polar scattering angle $\theta_{\text{pol}}$ and on the incoming proton momentum at the center of the analyzer. The efficiency [38] is also function of the analyzer thickness—the longer it gets, the higher the number of target nuclei to rescatter off. Two different parametrizations of the analyzing power are available in McNaughton's [37] paper for the momentum range of interest for this work, a low-energy and high-energy parametrization for kinetic energies at the center of the analyzer up to 450 MeV and 750 MeV respectively. The figure of merit [38] for a polarimeter can be characterized as:

$$F^2 = \int_{\theta_{\text{min}}}^{\theta_{\text{max}}} A_p(\theta_{\text{pol}})^2 \epsilon(\theta_{\text{pol}}) d\theta_{\text{pol}}. \quad (A4)$$

In our calculations, the efficiency and analyzing power are integrated over the range of 4–19 degrees in $\theta_{\text{pol}}$.

In Fig. 11 a scan of the analyzing power, efficiency and figure of merit as a function of proton kinetic energy at the center of a Carbon analyzer $T_{\text{carb}}$ is shown. While the efficiency increases with kinetic energy, the analyzing power is the largest around $T_{\text{carb}} = 200$ MeV and the figure of merit for the polarimeter peaks between 200 and 300 MeV.

In Fig. 12, the analyzing power and efficiency are shown as a function of the scattering angle $\theta_{\text{pol}}$ for three different proton energies. The efficiency is seen to drop with $\theta_{\text{pol}}$, while the general shape on the analyzing power varies with the proton energy.

The statistical error on the measured polarization is proportional to:

$$\delta_P \propto \frac{1}{F\sqrt{N_{\text{inc}}}}, \quad (A5)$$

with $N_{\text{inc}}$ denoting the number of incident protons.

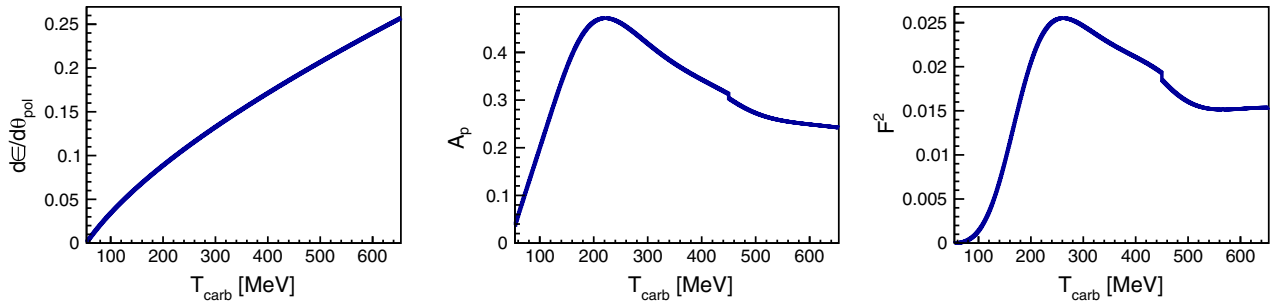

FIG. 11. A scan over differential efficiency (left), analyzing power (center), and $F^2$ (right) as a function of the kinetic energy in the center of the analyzer. The shown range corresponds to a proton momentum of 320–1290 MeV/c at the center of the analyzer.

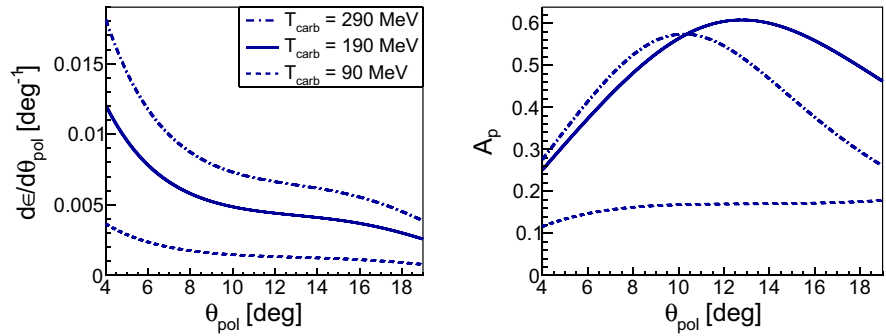

FIG. 12. The differential efficiency (left) and analyzing power (right) as a function of $\theta_{pol}$ for a proton with kinetic energies of 90, 190 and 290 MeV at the center of the carbon analyzer.